\documentclass[twocolumn,prb,superscriptaddress,aps,showpacs]{revtex4}

\usepackage{graphicx}
\usepackage{dcolumn}
\usepackage{amsmath}
\usepackage{tikz}
\usepackage{subfigure}
\begin{document}
\title{Spin-Orbit Exciton in a Honeycomb Lattice Magnet CoTiO$_3$: Revealing Link Between Rare Earth and Transition Metal Magnetism}
\author{Bo Yuan}
\affiliation{Department of Physics, University of Toronto, Toronto, Ontario, M5S 1A7, Canada}

\author{M. B. Stone}
\affiliation{Neutron Scattering Division, Oak Ridge National Laboratory, Oak Ridge, Tennessee 37831, USA}

\author{Guo-Jiun Shu}
\affiliation{Department of Materials and Mineral Resources Engineering, National Taipei University of Technology, Taipei 10608, Taiwan}
\affiliation{Institute of Mineral Resources Engineering, National Taipei University of Technology, Taipei 10608, Taiwan}
\affiliation{Taiwan Consortium of Emergent Crystalline Materials, Ministry of Science and Technology, Taipei 10622, Taiwan}

\author{F. C. Chou}
\affiliation{Center for Condensed Matter Sciences, National Taiwan University, Taipei, 10617 Taiwan}

\author{Xin Rao}
\affiliation{Department of Physics, Hefei National Laboratory for Physical Sciences at Microscale,and Key Laboratory of Strongly-Coupled Quantum Matter Physics (CAS), University of Science and Technology of China, Hefei, Anhui 230026, People's Republic of China}

\author{J. P. Clancy}
\affiliation{Department of Physics and Astronomy, McMaster University, Hamilton, ON L8S 4M1 Canada}

\author{Young-June Kim}
\affiliation{Department of Physics, University of Toronto, Toronto, Ontario, M5S 1A7, Canada}

\begin{abstract}
We carried out inelastic neutron scattering to study the spin-orbital (SO) exciton in a single crystal sample of CoTiO$_3$ as a function of temperature. CoTiO$_3$ is a honeycomb magnet with dominant XY-type magnetic interaction and an A-type antiferromagnetic order below $\mathrm{T_N} \approx 38$~K. We found that the SO exciton becomes softer, but acquires a larger bandwidth in the paramagnetic phase, compared to that in the magnetically ordered phase. Moreover, an additional mode is only observed in the intermediate temperature range, as the sample is warmed up above the lowest accessible temperature below $\mathrm{T_N}$. Such an unusual temperature dependence observed in this material suggests that its ground states (an $S_{\mathrm{eff}}=\frac{1}{2}$ doublet) and excited states multiplets are strongly coupled, and therefore cannot be treated independently, as often done in a pseudo-spin model. Our observations can be explained by a multi-level theory within random phase approximation that explicitly takes into account both the ground and excited multiplets. The success of our theory, which is originally developed to explain temperature dependence of magnetic excitations in the rare-earth magnets, highlight the similarity between the magnetic excitations in rare-earth systems and those in transition metal systems with strong spin orbit coupling.

\end{abstract}
\maketitle
\section{Introduction}
Recently, there has been a dramatic increase in research effort to understand the effects of spin-orbit coupling (SOC) in a magnetic material. Anisotropic magnetic interactions such as Dzyalloshinskii-Moriya and Kitaev interactions arising from strong SOC are responsible for many exotic magnetic states such as non-trivial magnetic order found in multiferroic compounds\cite{Kimura2007,Tokura2010} and quantum spin liquid phases in Kitaev materials \cite{Kitaev2006,Chaloupka2010,Singh2010,Plumb2014,trebst2017,Winter_2017,Hermanns2018, Takagi2019}. Although the details of interactions driving the magnetic behaviour of a  material can be complex with a myriad of energy scales, such as Coulomb interaction, crystalline electric field (CEF), and SOC, the essential magnetism is, remarkably, often not directly dependent on these intra-atomic energy scales. In particular, this happens when the exchange interaction between magnetic ions, $\mathcal{J}$, is much smaller than these intra-atomic energy scales. In such a  case, it is sufficient to ignore the coupling between the ground and excited multiplets and simply project $\mathcal{J}$ onto the ground states manifold. The magnetic ground states and the low energy excitations can therefore be captured using a model of interacting pseudo-spins $\mathrm{S}_\mathrm{eff}$ \cite{Jackeli2009,Liu2018}. Since the ground multiplet over which $\mathrm{S}_\mathrm{eff}$ is defined usually has an unquenched orbital angular momentum, pseudo-spin interactions are anisotropic, which is the most important feature common to all transition-metal magnetic materials with large SOC. So far, materials with a Kramers doublet ground state equivalent to a $\mathrm{S}_\mathrm{eff}=\frac{1}{2}$ pseudo-spin have attracted the most attention due to their simplicity\cite{BJKim2008,Rau2016}. The same concept can be nontheless applied to other systems with different ground state degeneracy by defining pseudo-spins of different magnitude, although the resultant effective model can be slightly more complex\cite{chen2010,chen2011}.

Although the pseudo-spin picture greatly simplifies the description of low energy physics in these systems, the underlying assumption that the coupling between ground and excited multiplets is weak fails in some cases. When the magnetic interaction is large, the excited multiplets, often referred to as a spin-orbit (SO) exciton, can become highly dispersive and couple strongly to the ground states. Although largely neglected in transition-metal magnetic materials, the need to consider excited multiplets has been well recognized in a different class of materials, the rare-earth magnetic systems. The overall energy scale is considerably reduced in rare-earth materials as the CEF splitting is much smaller than that in a transition-metal material due to highly localized f-orbitals. As a result, the energy splitting between the ground and excited multiplets is often comparable to magnetic interactions, which invalidates a simple pseudo-spin picture where these multiplets are considered to be decoupled. Instead, dynamics of the higher energy multiplets become strongly dependent on what happens in the low energy sector, and vice versa. Direct evidence for this is a strong renormalization of the higher energy SO excitations across magnetic ordering that has been observed in some rare-earth systems \cite{Morin_1980,Castets1982,Hennion1978,Hennion1979,Halg1986,Sablik1979}. These observations can be only explained through a multi-level theory\cite{Haley1972,Buyers1975} where ground and excited multiplets are considered simultaneously.

Only recently, the important roles played by the excited multiplets are beginning to be recognized in some transition-metal magnetic materials. Perhaps the most dramatic example is the so-called ‘excitonic’ magnetism in systems with a nominally non-magnetic $\mathrm{S}_\mathrm{eff}=0$ singlet ground state, first proposed for rare-earth intermetallics such as PrTl$_3$\cite{Andres1972} and more recently extended to heavy $d^4$ transition-metal ions\cite{Khaliulin2013}. When dispersion of the SO exciton is large enough to cross the ground state, magnetic order may be induced in these systems via condensation of the SO exciton. Other than this special case of a $\mathrm{S}_\mathrm{eff}=0$ magnet, a strongly dispersive SO exciton has also been observed in a number of iridates\cite{Kim2014,Kim2012} and cobaltates\citep{Buyers_1971,Holden_1971,Sarte2019} with a $\mathrm{S}_\mathrm{eff}=\frac{1}{2}$ ground state and a simple magnetic order, which raises important questions on the applicability of a simple psuedo-spin picture in these transition-metal systems.

We examined these questions by studying high-energy magnetic excitations around the SOC energy scale in a typical transition-metal magnet with strong SOC.
The transition-metal system we focus on in our study is CoTiO$_3$ with an ilmenite structure, where the magnetic Co$^{2+}$ ions form a honeycomb layer, ABC-stacked along the $c$ direction (See Fig.~\ref{fig1}(i)). Each Co$^{2+}$ residing in a trigonally distorted octahedra has an pseudo-spin $\mathrm{S}_\mathrm{eff}=\frac{1}{2}$ doublet ground state determined by a combination of SOC and CEF (See Ref.~\onlinecite{yuan2019} as well as below). Below $\mathrm{T_N}\approx38~\mathrm{K}$, the Co$^{2+}$ pseudo-spins are ordered ferromagnetically in the $ab$ plane and antiferromagnetically along $c$ \cite{Newnham1964}. A strong easy-plane magnetic anisotropy was inferred from the large $\frac{\chi_\parallel}{\chi_\perp}$ observed in the bulk magnetization data\cite{WATANABE1980} and also from the magnetic structure determined by neutron diffraction. Due to the simplicity of both the magnetic and crystal structure, CoTiO$_3$ was considered an ideal model system to study 3D XY-magnetism, prompting extensive studies to investigate the effects of doping by magnetic\cite{Toshiya1991Mndoped,Yasuo1986,Ito1982,NEWSAM1983,Harris1997ZPB,Harris1997} and non-magnetic ions\cite{TAYLOR2001Gedoped}. In particular, the solid-solution of CoTiO$_{3}$ and FeTiO$_{3}$, the latter of which is an Ising system, has been systematically studied to construct the full phase diagram of a mixed anisotropy system\cite{Yasuo1986,Ito1982,NEWSAM1983,Harris1997ZPB,Harris1997}. However, the first comprehensive inelastic neutron scattering (INS) measurements were carried out very recently \cite{yuan2019}, in which direct dynamical evidence in the magnon spectrum for strong XY exchange anisotropy in CoTiO$_3$ was provided. More importantly, the existence of Dirac magnon with linear crossing at the $\mathbf{K}$ point of the Brillouin zone in CoTiO$_3$ was discovered in the same study, which renewed interest in this material as a potential candidate system for studying topological magnons.

In this paper, we report INS study on the high energy SO exciton in CoTiO$_3$. Using a single crystal sample, we directly probed its momentum dependent excitation spectrum at various temperatures across $\mathrm{T_N}$. We observed a strong temperature dependence of the high energy SO exciton across $\mathrm{T_N}$. This strongly contradicts a simple pseudo-spin picture, which suggests that the excited multiplets are decoupled from the ground multiplet. Instead, it reminisces the behaviours observed in many rare-earth systems. The observed temperature dependence is well explained by a multi-level theory within random phase approximation (RPA) originally developed for the rare-earth systems\citep{Buyers1975}, which further highlights the strong similarity between the two classes of materials. Quantitatively however, we found a model using simple bilinear spin interaction underestimates the size of the SO exciton's bandwidth at all temperatures, suggesting potential presence of higher order spin interactions in this material.

\section{Experimental Details}
The same single crystal samples used in Ref.~\onlinecite{yuan2019} was used in this experiment. It is grown in two stages. First, single phase polycrystalline powder of CoTiO$_3$ was synthesized with the solid-state reaction method using raw materials of Co$_3$O$_4$ and TiO$_2$ of 4N purity. The stoichiometric mixture with molar ratio of Co:Ti=1:1 was wet milled using acetone for 12 hours with zirconia ball milling. The thoroughly mixed fine grain slurry was dried in argon flow first, and then reacted at 1200~$^\circ$C for 48 hours in Ar atmosphere. The final product was slowly cooled down to room temperature under the Ar gas flow. After the high purity powder is made, single crystals were grown with an Optical Floating Zone furnace (Crystal System Inc.) having 4$\times$1000~W halogen lamps as the heating source. The CoTiO$_3$ polycrystalline powder was ground and shaped into feed rods of $\sim$6~mm diameter and $\sim$100~mm long using an isostatic pressure of $\sim$100~MPa. The rods were sintered at 1250~$^\circ$C for 12 hours in Ar flow. Because of the congruent melt nature, the stoichiometric CoTiO$_3$ was used for both feed and seed rods. Successful growth was obtained using a pulling rate of 3 mm/h and 20 rpm rotation in opposite direction in an Ar flow rate of 50~ml/min. Other than a single crystal sample grown as described above, another powder sample of CoTiO$_3$ was also used in this experiment. It was synthesized by mixing stoichiometric amount of CoCO$_3$ and TiO$_2$ and kept in air at 1150$~^\circ$C for 72 hours, with intermediate grinding at every 24 hours interval.

The single crystal sample was aligned at McMaster Alignment Diffractometer (MAD) at the McMaster reactor before the inelastic neutron scattering (INS) experiment. INS measurement was carried out at the SEQUOIA time-of-flight spectrometer at SNS, ORNL. Measurements on single crystals were carried out using an incident neutron energy of $\mathrm{E_i}=50~\mathrm{meV}$. Two chopper settings were used: a high resolution (HR) setting using T0 and FC2 choppers rotating at a frequencies of 90~Hz and 360~Hz as well as a high flux (HF) setting using T0 and FC1 choppers rotating at a frequencies of 90~Hz and 180~Hz. Energy resolutions of $\sim1.4~\mathrm{meV}$ and $\sim2.9~\mathrm{meV}$ were achieved for the HR and HF settings, respectively. A high $\mathrm{E_i}=250~\mathrm{meV}$ measurement was carried out on powder to determine the crystal field levels. The incident energy was selected by rotating the T0 and FC1 choppers at frequencies of 120~Hz and 360~Hz, respectively, which gave an energy resolution of $\sim$20~meV at the elastic line. Temperatures used in the measurements were controlled by a closed cycle refrigerator.

\section{Experimental Results}
\begin{figure*}[tb]
	\centering
\includegraphics[width=1\textwidth]{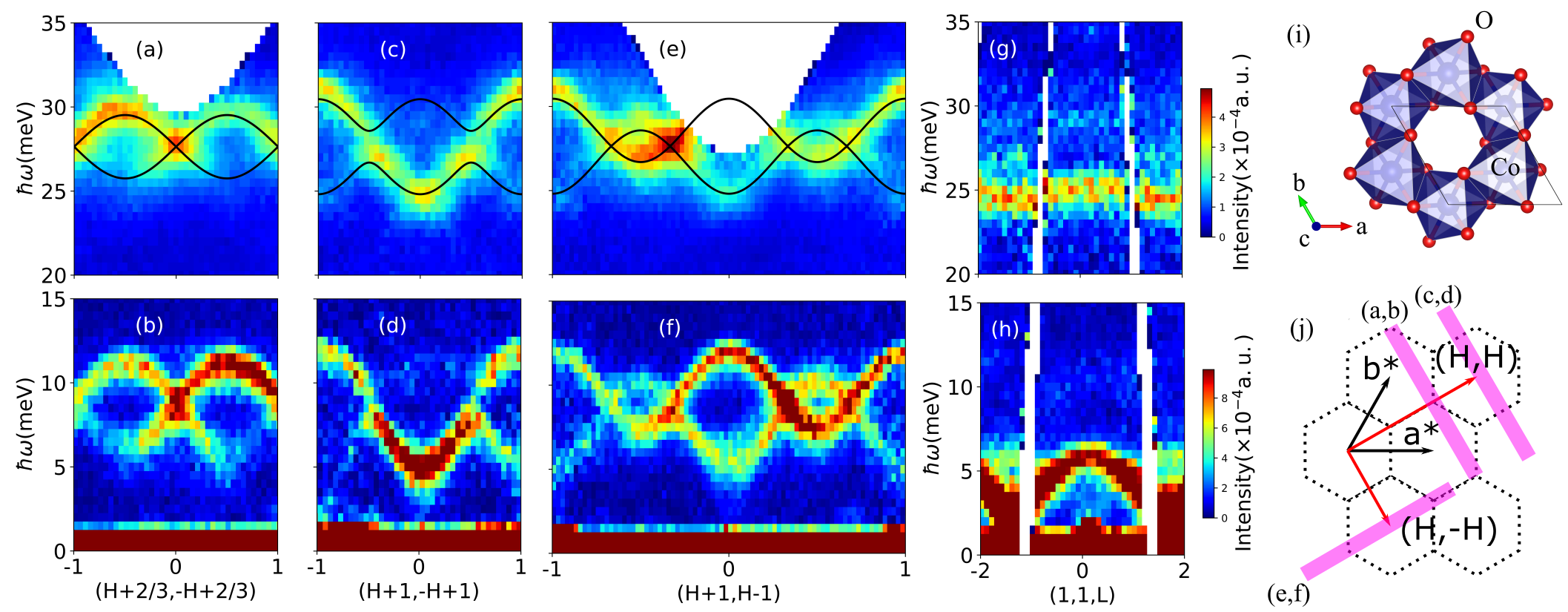}
	\caption{(a-h) Neutron intensity as a function of energy $\hbar\omega(\mathrm{meV})$ and momentum transfer detailing (top row) the high energy SO exciton and the (bottom row) low energy magnon in CoTiO$_3$. The data were obtained at the SEQUOIA time-of-flight spectrometer at $\mathrm{T=5~\mathrm{K}}$ using an incident energy of $\mathrm{E_i=50~\mathrm{meV}}$ and a high resolution chopper setting. Since the SO exciton in (g) shows almost no dispersion along L, data in (a,c,e) have been integrated along L to improve data quality.  Solid black lines are fit to a nearest neighbour tight binding model on a honeycomb lattice as described in the text. On the other hand, the magnons are considerably more dispersive along L as shown in (h). Neutron spectra in (e,d,f) are obtained for fixed L=0.5 by integrating neutron intensity over a small range along L=$\left[0.3,0.7\right]$. Data in the top/bottom row are shown on the intensity scale to the right of panel (g)/(h). (i) Structure of CoTiO$_3$ in a Co honeycomb layer together with the surrounding oxygen octahedra (j) 2D projection of the Brillouin zones showing directions of momentum transfers within the honeycomb plane (thick purple lines) in (a-f).}
	\label{fig1}	
\end{figure*}

High resolution INS data at $\mathrm{T=5~\mathrm{K}}$ are shown in Fig.~\ref{fig1} as a function of energy $\hbar\omega$ (vertical axis) and momentum transfer (horizontal axis). Directions of momentum transfers in Fig.~\ref{fig1}a-f are shown as thick pink lines through the 2D projection of Brillouin zones in Fig.~\ref{fig1}j. The INS spectra shown in Fig.~\ref{fig1} consist of two bands of magnetic excitations: a low energy band below 15~meV and a high energy band above 20~meV. In this paper, we focus on the high energy magnetic excitations occurring above 20~meV (top row of Fig.~\ref{fig1}). The low energy magnetic excitations correspond to magnons in the magnetically ordered phase of CoTiO$_3$ (bottom row of Fig.~\ref{fig1}) and have been studied in detail in Ref.~\onlinecite{yuan2019}. As we will show later, the high energy magnetic excitation with a clear dispersion between 25~meV and 30~meV is attributed to a SO exciton of trigonally distorted Co$^{2+}$ ions. This SO exciton, already present at 5~K, will henceforth be referred to as mode $\mathbf{B}$ (the magnon will be referred to as mode $\mathbf{A}$).

As shown in Fig.~\ref{fig1}(a,c,e), the SO exciton is strongly dispersive within the honeycomb plane but has an order of magnitude smaller dispersion ($\sim0.5~\mathrm{meV}$) in the out of plane direction (Fig.~\ref{fig1}g), suggesting its quasi-2D nature. Phenomenologically, its in-plane dispersion can be modelled as
\begin{align}
\omega(\vec{q})=\pm t\left|\sum_{\vec{d}}\exp(-i\vec{q}\cdot\vec{d})\right|+\Delta,\
\label{TB}
\end{align}
which is the same as the quasi-particle dispersion of a tight-binding model on honeycomb lattice. In this expression, $t$ and $\Delta$ denote the nearest neighbour hopping and an overall energy shift\cite{Hyart2018} of the SO exciton, respectively, and $\vec{d}$ denotes a vector connecting an atom to its three nearest neighbours. Two branches predicted by the model, which are also observed in our data, correspond to quasi-particle excitations on the two sub-lattices of a honeycomb lattice. As shown by the solid lines in Fig.~\ref{fig1}a-c, this simple phenomenological model gives an excellent description of the 5~K data, with $t_B(\mathrm{5~K})=0.9(2)~\mathrm{meV}$ and $\Delta_B(\mathrm{5~K})=27.6(2)~\mathrm{meV}$ (The subscript $B$ denotes the relevant parameters for mode $\mathbf{B}$). As we discuss below, microscopic origin of the hopping $t$ comes from magnetic interactions between Co$^{2+}$ ions. The observation that only the $\it{nearest\,neighbour}$ hopping is required to describe the data indicates that only the nearest neighbour magnetic interaction is important in CoTiO$_3$, a conclusion also reached in our previous study of low energy magnons\cite{yuan2019}.

\begin{figure*}[tb]
	\centering
\includegraphics[width=1\textwidth]{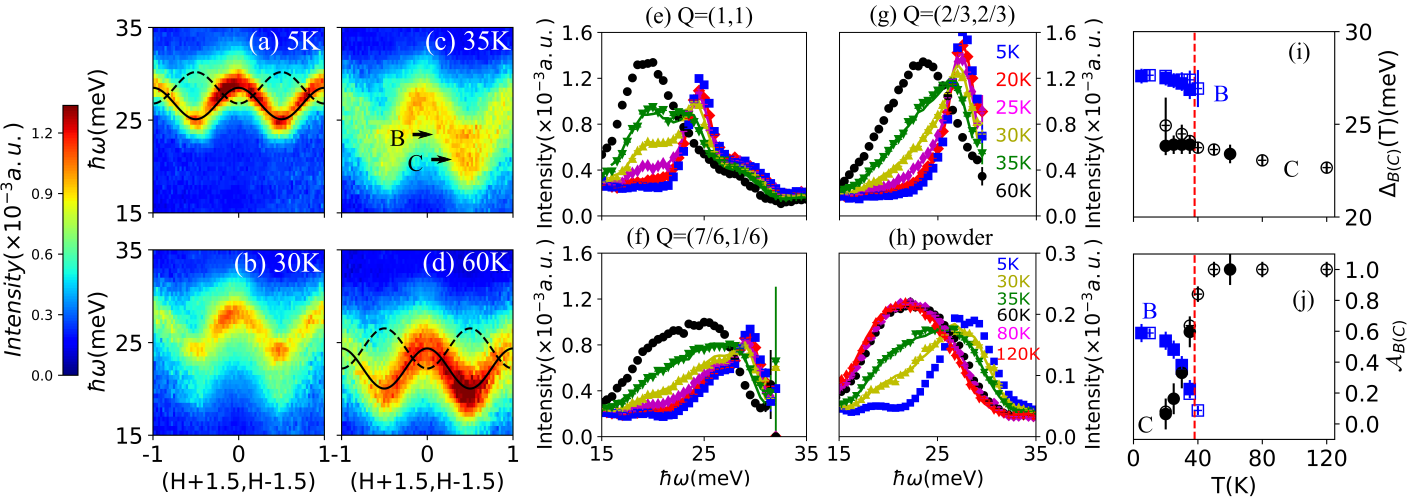}
	\caption{(a-d) Temperature dependence of neutron spectra along (H+1.5,H-1.5). The $\mathbf{B}$ and $\mathbf{C}$ modes are marked by black arrows in (c) (e-g): Constant-Q cut of the single crystal high flux INS data at (e) $\mathbf{Q}=(1,1)$ (f) $\mathbf{Q}=(\frac{2}{3},\frac{2}{3})$ and (g) $\mathbf{Q}=(\frac{7}{6},\frac{1}{6})$  at temperatures from 5~K to 60~K. (h) Constant-Q cut of the powder-averaged high resolution INS data from 5~K to 120~K. The `powder'-averaged INS data here is obtained by rotating the single crystal over 360$^\circ$ and averaging over all sample orientations. Intensity with $|\mathbf{Q}|=\left[1.0~\AA^{-1},4.0~\AA^{-1}\right]$ was then integrated and plotted as a function of $\hbar\omega$. Solid lines in (e-h) are fit to the cuts from 20~K-35~K using a linear combination of data at 5~K and 60~K as described in the text. (i,j) Temperature dependence of (i) average energy and (j) relative intensity of modes $\mathbf{B}$ (blue) and $\mathbf{C}$ (black). Energy of the mode $\mathbf{B}$($\mathbf{C}$) or $\Delta_{B(C)}(\mathrm{T})$, is obtained by subtracting a temperature-dependent shift $\delta_{B(C)}$ from the energy of mode $\mathbf{B}$ ($\mathbf{C}$) at 5~K (60~K). As described in the text, the energy shift $\delta_{B(C)}$ and relative intensity $\mathcal{A}_{B(C)}$ of mode $\mathbf{B}$($\mathbf{C}$) are obtained from fitting to constant-Q cuts at each temperature using Eq.~\eqref{intensity}. Filled and open symbols are obtained from fitting the single crystal and powder averaged data, respectively. Position of Neel temperature $\mathrm{T_N=38K}$ has been denoted by vertical red dashed line.}
	\label{fig2}	
\end{figure*}

Very interestingly, a second mode (referred to as $\mathbf{C}$) with almost the same dispersion emerges at an energy slightly below $\mathbf{B}$ as temperature increases. This is illustrated in the INS data in Fig.~\ref{fig2}, where we show the neutron spectra along (H+1.5,H-1.5) at different temperatures from 5~K to 60~K. At 5K (Fig.~\ref{fig2}a), only mode $\mathbf{B}$ centered at $\sim27~\mathrm{meV}$ is visible with a W-shaped dispersion along (H+1.5,H-1.5). Only one branch of Eq.~\eqref{TB} is visible in Fig.~\ref{fig2}a (shown by the black solid line) while the other (black dashed line) is suppressed due to small dynamical structure factor in this Brillouin zone. (This is clear by comparing Fig.~\ref{fig2}a with Fig.~\ref{fig1}e, which shows SO exciton along an equivalent direction in a different Brillouin zone where both branches are clearly visible.) At 30~K and 35~K, intensity of $\mathbf{B}$ decreases while another W-shaped mode,$\mathbf{C}$, starts to gain intensity. Coexistence of $\mathbf{B}$ and $\mathbf{C}$ is most clearly seen in the 35~K data (Fig.~\ref{fig2}c), where the two modes have been indicated with the horizontal black arrows. At 60~K ($>\mathrm{T_N}$), only one mode is visible which we continue to label as $\mathbf{C}$. As shown in Fig.~\ref{fig2}d, mode $\mathbf{C}$ at 60~K in the paramagnetic phase appears to be damped and occurs at a lower energy $\Delta_C(\mathrm{60~K})=23.3(2)~\mathrm{meV}$ compared to $\mathbf{B}$. Fitting to the same phenomenological model given by Eq.~\eqref{TB} gives a larger hopping $t_C(\mathrm{60~K})=1.2(2)~\mathrm{meV}$ at 60~K, suggesting an increase in the bandwidth of the SO exciton for $\mathrm{T~>~T_N}$.

\begin{figure}[tb]
\includegraphics[width=0.4\textwidth]{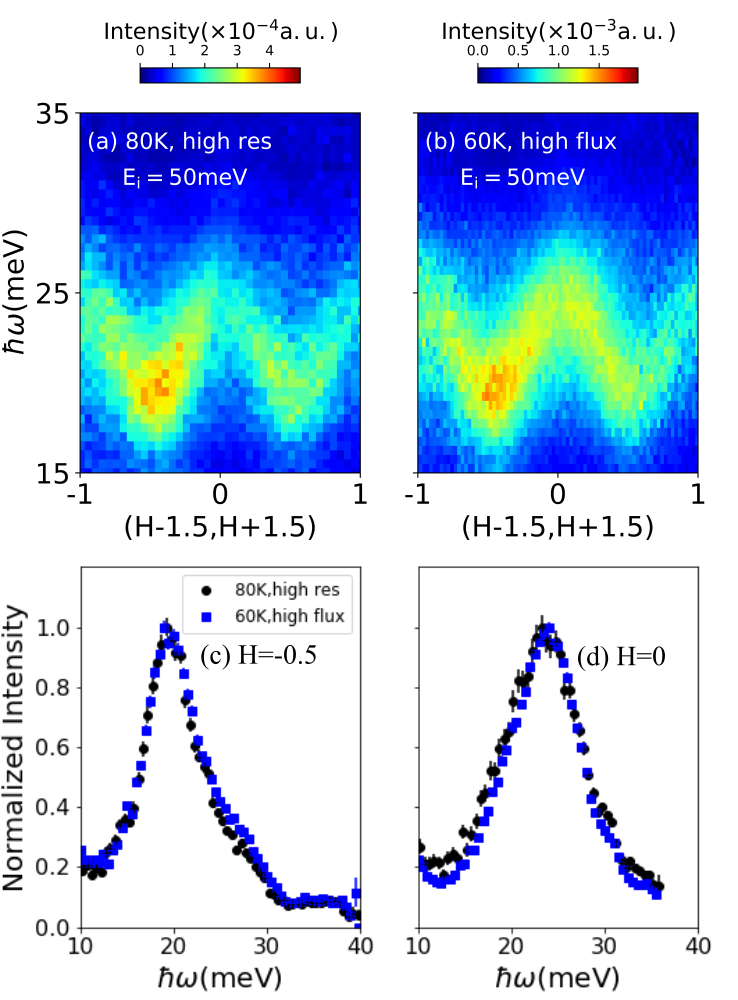}
\caption{Neutron spectra (a) at 80~K obtained with $\mathrm{E_i=50~meV}$ and the high resolution chopper setting (b) at 60~K obtained with the same incident energy but using the high flux chopper setting. (c,d) Comparison between constant-Q cuts of (a) and (b) at (c) H=-0.5 and (d) H=0. (c) and (d) have been normalized with respect to the peak intensity.}
\label{supp1}
\end{figure}

To study the SO exciton at intermediate temperatures more carefully, we make constant-Q cuts of the INS data. Representative cuts are shown in Fig.~\ref{fig2}e-g for different $\mathbf{Q}$ positions. At 5~K, the constant-Q cut in Fig.~\ref{fig2}e with $\mathbf{Q}=(1,1)$ consists of two peaks at $\sim25~\mathrm{meV}$ and $\sim30~\mathrm{meV}$, both corresponding to the two branches of $\mathbf{B}$. At higher temperatures (but still below $\mathrm{T_N}$, for example at 30~K), the constant Q cut acquires a complex line shape with additional spectral weight appearing below the double peaks due to $\mathbf{B}$. Constant-Q cuts at other $\mathbf{Q}$'s shown in Fig.~\ref{fig2}f,g show a similar asymmetrical peak profile at intermediate temperatures, where additional spectral weight appears due to development of a new mode. These observations are further corroborated by the following more quantitative analysis. We find constant-Q cuts at intermediate temperatures $\mathrm{5~K<T<60~K}$ to be fit very well by a simple linear combination of data at two limiting temperatures, 5~K and 60~K, where only $\mathbf{B}$ and $\mathbf{C}$ modes have nonzero intensities, respectively:
\begin{align}
I_T(\hbar\omega)=\mathcal{A}_B I_{\mathrm{5K}}(\hbar\omega-\delta_B)+\mathcal{A}_C I_\mathrm{60K}(\hbar\omega-\delta_C).\label{intensity}
\end{align}
In Eq.~\eqref{intensity}, $I_\mathrm{T}(\hbar\omega)$ is the intensity as a function of energy in a constant-Q cut at a temperature $\mathrm{T}$. The first (second) term denotes the empirical lineshape of  the 5~K (60~K) data with an overall scaling of intensity $\mathcal{A}_B$ ($\mathcal{A}_C$) and an overall shift in energy $\delta_B$($\delta_C$). Most remarkably, this simple expression using the $\it{same}$ sets of $\{\mathcal{A}_B,\mathcal{A}_C,\delta_B,\delta_C\}$ are found to simultaneously fit the data at $\it{all}$ Q-positions shown in Fig.~\ref{fig2}e-g. This provides a robust way to extract the positions and relative intensities of modes $\mathbf{B}$ and $\mathbf{C}$, which are very close in energy at the intermediate temperatures. Moreover, since $\{\mathcal{A}_B,\mathcal{A}_C,\delta_B,\delta_C\}$ does not depend on $\mathbf{Q}$, the powder averaged data can also be fit using Eq.~\eqref{intensity} as in Fig.~\ref{fig2}h, providing additional data points (both $\mathrm{T<T_N}$ and $\mathrm{T>T_N}$) supporting our conclusion that the lineshape at an intermediate temperatures is well described by a simple linear combination of the 5~K and 60~K data. In addition, constant-Q cuts of the powder averaged data in Fig.~\ref{fig2}h above $\mathrm{T_N}$ at 80~K and 120~K are identical to that at 60~K, indicating that the SO exciton is unchanged across a wide range of temperatures with $\mathrm{T>T_N}$. This confirms that the observed temperature dependence in Fig.~\ref{fig2}a-d from 5~K to 60~K is associated with the onset of magnetic order. This is also supported by our high resolution single crystal measurement at 80~K (Fig.~\ref{supp1}) showing only one broad dispersive SO exciton mode identical to the high flux data at 60~K in Fig.~\ref{fig2}d.

The fitting results are summarized in Fig.~\ref{fig2}i and \ref{fig2}j. The results are not shown for mode $\mathbf{B}$ ($\mathbf{C}$) at $\mathrm{T>40~K}$ ( $\mathrm{T<20~K}$) where the intensity is negligible. At a temperature, $\mathrm{T}$, energy of mode $\mathbf{B}$ ($\mathbf{C}$) is obtained by subtracting the temperature -dependent shift $\mathrm{\delta_B}$ ($\mathrm{\delta_C}$) from energy of the mode at 5~K (60~K) and are given by $\mathrm{\Delta_B}(\mathrm{T})=\mathrm{\Delta_B}(\mathrm{5~K})-\mathrm{\delta_B}(\mathrm{T})$ ($\mathrm{\Delta_C}(\mathrm{T})=\mathrm{\Delta_C}(\mathrm{60~K})-\mathrm{\delta_C}(\mathrm{T})$). (Explicit temperature dependence is given in the bracket of each quantity for clarity.) As shown in Fig.~\ref{fig2}i, mode $\mathbf{B}$ slightly softens while energy of $\mathbf{C}$ is relatively unchanged as temperature approaches $\mathrm{T_N}$. In Fig.~\ref{fig2}j, we show temperature dependence of intensities of $\mathbf{B}$ and $\mathbf{C}$. Intensity of $\mathbf{B}$ is suppressed as that of $\mathbf{C}$ increases. The most dramatic change occurs near $\mathrm{T_N}$ denoted by the vertical red dashed line.

\section{Multi-level multiplet Theory}

The observed temperature dependence is surprising. First as temperature increases, it is the lower energy mode that gains intensity ($\mathbf{C}$), while the higher energy mode ($\mathbf{B}$) loses intensity. Second, there is an increase in the SO exciton's bandwidth above $\mathrm{T_N}$. The fact that the SO exciton is dispersive at all in the paramagnetic phase is quite surprising as it is hard to imagine the coherent propagation of a magnetic exciton in the absence of magnetic order. However, as mentioned in the introduction, most of our observations can be naturally explained when the ground and excited multiplets are considered simultaneously.

\subsection{Single-Ion Hamiltonian}
\begin{figure}[tb]
\includegraphics[width=0.4\textwidth]{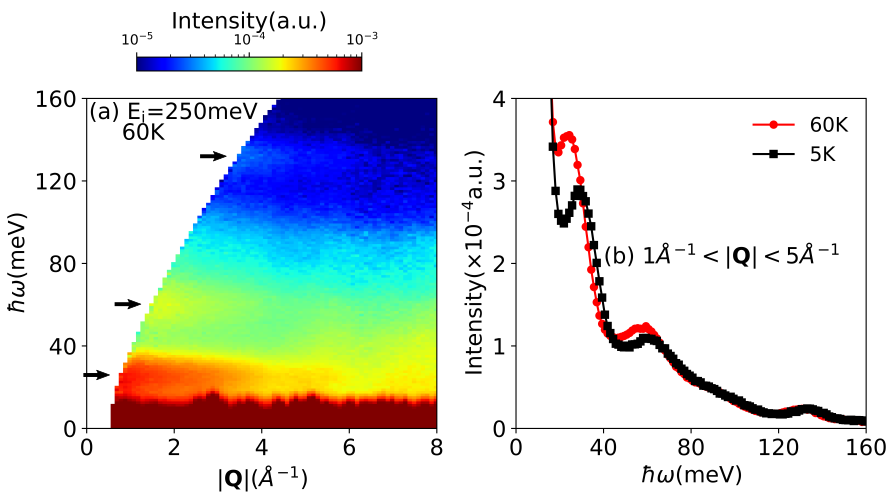}
\caption{(a) Powder averaged INS spectrum at 60~K obtained using an incident energy $\mathrm{E_i=250~meV}$. Unlike Fig.~\ref{fig2}h, the measurement here was carried out on an actual powder sample. A logarithmic intensity scale is used. Three CEF transitions have been marked by horizontal arrows. (b) Constant-Q cut of the powder-averaged INS data ($\mathrm{E_i}$=250~meV) at 60~K and 5~K. Intensity with $|\mathbf{Q}|=\left[1.0~\AA^{-1},5.0~\AA^{-1}\right]$ was integrated and plotted as a function of $\hbar\omega$.}
\label{highEi}
\end{figure}

Many features of our data can already be understood by examining what happens to a single Co$^{2+}$ ion. In trigonally distorted CoTiO$_3$, the single ion Hamiltonian $H_1$ is given by
\begin{align}
H_1=\Delta_\mathrm{trig}L_z^2+\lambda \vec{S}\cdot\vec{L}+h_0\left\langle S_x\right\rangle S_x.\label{SI}
\end{align}
where $L=1$ and $S=\frac{3}{2}$ describes the orbital and spin angular momenta of a Co$^{2+}$ ion in the high spin state. In this expression, $\Delta_\mathrm{trig}$ and $\lambda$ gives the trigonal distortion and spin-orbit coupling in CoTiO$_3$. The third term proportional to $h_0$ gives the molecular field on a spin due to exchange interactions with the surrounding ions in the magnetically ordered phase. The dominant nearest neighbour spin interaction is assumed to take on a simple Heisenberg form, $\mathcal{J}\vec{S}_i\cdot\vec{S}_j$, therefore $h_0=3\mathcal{J}$. \footnote{It is important to note that $\mathcal{J}$ refers to interaction between actual spins, NOT pseudo-spins. When projected onto the ground doublet, even $\mathcal{J}$ of a Heisenberg form considered here leads to strong XY-exchange anisotropy. For further details, see Ref.~\onlinecite{yuan2019}} In Eq.~\eqref{SI}, $\Delta_\mathrm{trig}$ and $\lambda$ are the dominant energy scales. They are determined from a high energy ($\mathrm{E_i}$=250~meV) INS measurement shown in Fig.~\ref{highEi}. A constant-Q cut clearly reveal three peaks occuring at 23(2)~meV, 58(2)~meV and 132(2)~meV. Since the measurement was carried out at 60~K($>\mathrm{T_N}$) in the paramagnetic phase, the observed peaks are transitions between different energy levels of Eq.~\ref{SI} for $h_0=0$. We can assume only the ground state is populated at temperature of the measurement (60~K) and the observed transitions correspond to those from the ground state to excited states. The three transition energies uniquely determine $\Delta_\mathrm{trig}$ and $\lambda$ to be 57(6)~meV and 26(1)~meV, respectively. Value of $\lambda$ obtained here is consistent with other cobalt oxides \citep{Holden_1971}. In our previous work \citep{yuan2019}, a slightly different set of $\Delta_\mathrm{trig}$ and $\lambda$ were obtained as they were (incorrectly) determined from a CEF measurement at T=5~K ($<\mathrm{T_N}$). A constant Q cut at 5~K is also shown in Fig.~\ref{highEi}b for comparison. Energies of the CEF transitions at 5~K are slightly shifted compared to their values at 60~K due to a nonzero $h_0$.

Using $\Delta_\mathrm{trig}$ and $\lambda$ obtained above, we find that the 12 states of a Co$^{2+}$ ion (L=1,S=3/2) break into 6 Kramers doublets. Since the trigonal distortion breaks the full rotational symmetry of a free ion, the total angular momentum $J$ is no longer a good quantum number. However, Eq.~\ref{SI} still possesses rotational symmetry around the $z$ axis when $h_0=0$, the six Kramers doublets can therefore be labelled by $z$ component of the total angular momentum $J_z$. Importantly, wave-functions of the two sets of doublet with the lowest energies studied in our experiment are given by
\footnotesize\begin{align}
\begin{split}
\left|J_z=\pm \frac{1}{2}\right\rangle = -&0.27\left|J=\frac{5}{2},J_z=\pm \frac{1}{2}\right\rangle\\
\mp&0.18\left|J=\frac{3}{2},J_z=\pm \frac{1}{2}\right\rangle+0.95\left|J=\frac{1}{2},J_z=\pm \frac{1}{2}\right\rangle.
\end{split}
\label{W1}
\end{align}\normalsize

and

\footnotesize\begin{align}
\left|J_z=\pm \frac{3}{2}\right\rangle = 0.31\left|J=\frac{5}{2},J_z=\pm \frac{3}{2}\right\rangle \pm 0.95\left|J=\frac{3}{2},J_z=\pm \frac{3}{2}\right\rangle.
\label{W2}
\end{align}\normalsize

\begin{figure}[tb]
	\centering
\includegraphics[width=0.4\textwidth]{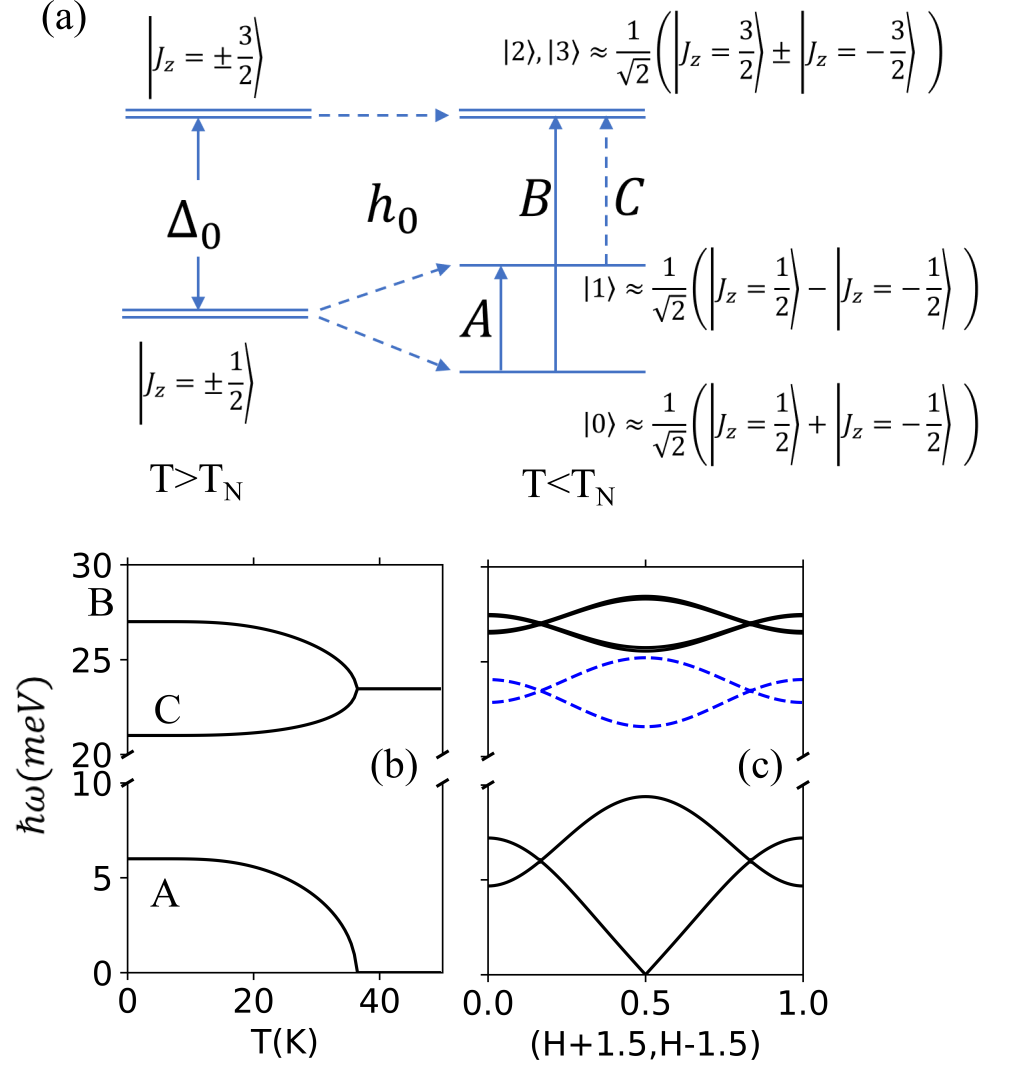}
	\caption{ (a) Schematic energy level diagram of the single ion model showing only the ground and first excited doublets. $h_0=0$ and $h_0\neq0$ limit are shown on the left and right, which represent the paramagnetic ($\mathrm{T>T_N}$) and magnetically ordered phases ($\mathrm{T<T_N}$), respectively. The approximate wave-functions are given besides the states. (b) Temperature dependence of energies of the transitions $\mathbf{A}$,$\mathbf{B}$ and $\mathbf{C}$ calculated self-consistently using the single ion model. (c) Dispersion of magnetic excitations at T=0 (black solid line) and $\mathrm{T>T_N}$ (blue dashed line). Both (b) and (c) are calculated with a $\Delta_0=23.5~\mathrm{meV}$ and $h_0=3\mathcal{J}=-3~\mathrm{meV}$. At T=0, both magnon and SO exciton are present. Only SO exciton is present at $\mathrm{T>T_N}$.}
	\label{fig3}	
\end{figure}

Since $h_0$ is much smaller than the splitting between any two doublets, we expect mixing induced by $h_0$ between the lowest two doublets and the rest of the states to be negligible. We therefore restrict our attention to the lowest two doublets probed in our experiment.The full Hamiltonian projected onto the lowest two doublets  (denoted by $\tilde{H}_1$) takes the following form

\begin{align}
\tilde{H}_1=\Delta_0\left(\left|\frac{3}{2}\right\rangle \left\langle \frac{3}{2}\right|+ \left|-\frac{3}{2}\right\rangle \left\langle-\frac{3}{2}\right|\right)+h_0\left\langle \tilde{S}_x \right\rangle \tilde{S}_x.
\label{SI1}
\end{align}
where $\tilde{S}_x$ is the projection of spin operator, $S_x$, for $S=\frac{3}{2}$ onto the lowest two doublets. Since the wave-functions given in Eq.~\eqref{W1} and Eq.~\eqref{W2} are eigenstates of $\Delta_\mathrm{trig}L_z^2+\lambda \vec{S}\cdot\vec{L}$, its projection onto these states gives the simple diagonal term in Eq.~\eqref{SI1} where $\Delta_0$ denotes the splitting between the two doublets. The energy level diagram of the lowest two doublets with and without $h_0$ are given in Fig.~\ref{fig3}a. With a non-zero $h_0$, degeneracy of the ground state doublet is broken while that of the first excited doublet approximately remains, as a molecular field term linear in spin operator $S_x$ does not directly couple the two states $|\pm\frac{3}{2}\rangle$. Fig.~\ref{fig3}a naturally accounts for temperature dependence of magnetic excitations in CoTiO$_3$. At $\mathrm{T=0}$, only the ground state is occupied, giving rise to two transitions: $\mathbf{A}$ and $\mathbf{B}$ (shown by vertical arrows with solid lines), corresponding to magnon and the SO exciton shown in our data. As the higher energy state of the ground doublet is thermally populated at higher temperatures $0<\mathrm{T}<\mathrm{T_N}$, another transition $\mathbf{C}$ (dashed arrow) appears below $\mathbf{B}$. Lastly, when $\mathrm{T>T_N}$, only one transition at an energy $\Delta_0$ is visible.

Quantitatively, the parameters $\Delta_0$ and $h_0$ of the projected single ion Hamiltonian can be determined from energies of the transitions $\mathbf{A}$ and $\mathbf{B}$, corresponding to magnon and SO exciton at T=0. Neglecting their dispersions, their energies are taken to be centers of the dispersions and are estimated to be 6~meV and 27~meV respectively from our data (Fig.~\ref{fig1}). This gives $\Delta_0=23.5~\mathrm{meV}$ and $h_0=-3~\mathrm{meV}$.

Having determined all parameters of the projected single-ion Hamiltonian, we can now calculate the the temperature dependence of transition energies $\mathbf{A}$, $\mathbf{B}$ and $\mathbf{C}$. For that, Eq.~\ref{SI1} needs to be diagonalized at each temperature. The unknown average $\left\langle\tilde{S}_x\right\rangle$ can be determined from
\begin{align}
\left\langle \tilde{S}_x \right\rangle=\sum_{m} \left\langle m\left| \tilde{S}_x\right|m \right\rangle \exp(-E_m/k_B T).
\label{temp}
\end{align}
where the sum runs over all four states shown in Fig.~\ref{fig3}a. Since both energies and wave-functions of these states depend on the thermal average $\left\langle\tilde{S}_x\right\rangle$, Eq.~\eqref{temp} has to be solved self-consistently at each temperature. Temperature dependence of transition energies found this way are given in Fig.~\ref{fig3}b. Two features are noteworthy. First, as temperature increases from 0 to $\mathrm{T>T_N}$, the transition energy from the ground state to the excited doublet decreases from 27~meV (energy of transition $\mathbf{B}$) to 23.5~meV ($\Delta_0$), in reasonable agreement with our data. Second, $h_0=-3~\mathrm{meV}$ predicts a $\mathrm{T_N}$ of 36~K, in good agreement with the experimental value of 38~K.

\subsection{Dispersion of SO Exciton}
To obtain the dispersion of SO exciton in the paramagnetic phase ($\mathrm{T>T_N}$), we apply the theory of generalized susceptibility within random phase approximation (RPA) developed by Buyers\cite{Buyers1975}, which was originally used to model the temperature dependence of magnetic excitations in rare-earth systems. Within this theory, the magnetic Hamiltonian $H$ is split into two parts, a single ion part $H_1$ consisting of intra-atomic interactions including SOC, CEF and molecular field due to surrounding ions, as well as an interaction part $H_2$, consisting of two-ion interactions. One first diagonalizes $H_1$ to extract the single ion eigenstates $| m\rangle$ with energy $E_m$, which is done in the last section. Magnetic excitations probed in a neutron scattering experiment correspond roughly to dipole allowed transitions between these single-ion levels.  One then includes $H_2$ to obtain the dispersions, or the q-dependent energy $\hbar\omega_{mn}(\vec{q})$, for an allowed transition from state $| m\rangle$ to $| n\rangle$. $\hbar\omega_{mn}(\vec{q})$ is given by poles in the Green's function $G^{\alpha\beta}(\vec{q},\omega)$, which is the fourier transform of $G^{\alpha\beta}(ij,t)=-i\theta(t)\left\langle \left[ S^\alpha_i(t),S^\beta_j \right] \right\rangle$, where $\theta(t)$ is the step function, square and angle brackets denote the commutator and thermal averaging respectively. The Fourier transform of the Green's function follows an equation of motion given by
\begin{align}
\omega G(S^\alpha_i,S^\beta_j,\omega)=\left\langle\left[ S^\alpha_i,S^\beta_j \right]\right\rangle+G(\left[S^\alpha_i,H\right],S^\beta_j,\omega).\label{Green}
\end{align}

To take into account all possible transitions, instead of working directly with spin operators, one decomposes the spin operator into boson operators $C_m^\dagger$'s which create the state $| m\rangle$, $S^\alpha_i=\sum_{mn} S^\alpha_{m,n} C^\dagger_{m,i} C_{n,i}$, where $S^\alpha_{m,n}=\left\langle m|S^\alpha|n \right\rangle$. The importance of this step is clear by noting that the operator $C_m^\dagger C_n$ induces a transition from $| m\rangle$ to $| n\rangle$. The desired dispersion relation $\hbar\omega_{mn}(\vec{q})$ is therefore directly contained in the dynamics of the operator $C_m^\dagger C_n$ described by the equation of motion Eq.~\eqref{Green}. When evaluating the commutator $\left[S_i^\alpha,H\right]$, one encounters a four boson term where one approximates using RPA as $C_m^\dagger(i) C_s(i) C_p^\dagger(j) C_q(j)\rightarrow f_m(i) \delta_{ms} C_p^\dagger(j) C_q(j)+f_p(j) \delta_{pq}C_m^\dagger(i) C_s(i)$. The parameters $f_m, f_p$ denote the thermal population of states $m,p$ in the single ion model. After this decomposition, one gets a set of linear equations coupling different components of the Green's function that can be subsequently solved.

Although this approach applies equally well to the magnetically ordered phase ($\mathrm{T<T_N}$), the loss of rotational symmetry around the $z$ axis when a molecular field is present greatly complicates the calculation. Instead, we apply a pseudo-boson approach\cite{Buyers_1971, Holden_1971} to obtain the dispersion of SO exciton at $\mathrm{T=0}$. (Equivalence between the pseudo-boson and generalized susceptibility approaches at $\mathrm{T=0}$ is shown in Ref.~\onlinecite{Buyers1975})

Details of the calculation are given in the Appendix, dispersion of SO exciton at $\mathrm{T=0}$ and $\mathrm{T>T_N}$ is shown in Fig.~\ref{fig3}. Notably, the theory correctly predicts a larger bandwidth for the SO exciton in the paramagnetic phase by $\sim1.3$ times, consistent with our data in Fig.~\ref{fig2}a-d.

\section{Discussions}
Our theory correctly predicts a larger bandwidth for the SO exciton at $\mathrm{T>T_N}$ (by $\sim$1.3 times) than at T=0 as observed experimentally in Fig.~\ref{fig2}a-d.
To gain an intuitive understanding of why this happens, we treat an excited state on a Co$^{2+}$ ion as an excitonic quasi-particle and study its motion under the exchange interaction $\mathcal{J}\vec{S}_i\cdot\vec{S}_j$. As an example, we consider the motion of a local exciton with wave-function $|2\rangle$ (defined in Fig.~\ref{fig3}a) that initially resides on site $j$ in the ordered phase. At T=0, this exciton moves in a uniform background of $|0\rangle$ states. It can hop to a neighbouring site $i$ (initially in its ground state $|0\rangle$) if wave-functions on the two sites can be exchanged. The effective hopping of the exciton is therefore given by the matrix element $t_\mathrm{T=0}=\mathcal{J}\left\langle 0,2\left|\vec{S}_i\cdot\vec{S}_j\right| 2,0\right\rangle=\mathcal{J}\left|\left\langle 0\left|\vec{S}\right|2\right\rangle\right|^2$, where $|0,2\rangle=|0\rangle_i\bigotimes |2\rangle_j$ denotes the two-ion state where site $i,j$ are occupied by $|0\rangle$ and $|2\rangle$, respectively. On the other hand, each site is populated by the two states in the ground doublet $\left|\pm\frac{1}{2}\right\rangle$ with equal probability at $\mathrm{T>T_N}$. Within a random phase approximation, the effective hopping of an excited state $\left|\frac{3}{2}\right\rangle$ is given by the thermal average $t_\mathrm{T>T_N}=\frac{1}{2}\mathcal{J}\left(\left|\left\langle \frac{1}{2}\left|\vec{S}\right|\frac{3}{2}\right\rangle\right|^2+\left|\left\langle -\frac{1}{2}\left|\vec{S}\right|\frac{3}{2}\right\rangle\right|^2\right)$ in the disordered phase. Using the single-ion wave-functions in the ordered and disordered phase, the effective hoppings are $t_\mathrm{T=0}=0.45\mathcal{J}$ and $t_\mathrm{T>T_N}=0.61\mathcal{J}$ respectively. Using $\mathcal{J}=\mathrm{-1~meV}$ and the fact that bandwidth is $6t$ for a nearest neighbour tight-binding model on a honeycomb lattice, bandwidths are estimated to be $\sim$2.7~meV and $\sim$3.6~meV for $\mathrm{T=0}$ and $\mathrm{T>T_N}$, in agreement with the treatment using Buyers' theory. From this simple argument, we can understand the change in bandwidth of the SO exciton as a change in effective hopping matrix element due to a change in the single ion wave-functions. More specifically, we argue that any difference in effective hopping in the ordered phase compared to the paramagnetic phase must be due to a small admixture of the ground and excited doublets. To see this, we neglect any mixing between the two doublets and diagonalize the molecular field term along $x$: $h_0\left\langle \tilde{S}_x\right\rangle \tilde{S}_x$ within each doublet in the ordered phase. As in Fig.~\ref{fig3}a, approximate wave-functions of the ground states doublet are given by $|0\rangle, |1\rangle\approx\frac{1}{\sqrt{2}}(\left|\frac{1}{2}\right\rangle\pm\left|-\frac{1}{2}\right\rangle)$; those of the excited states doublet are given by $|2\rangle, |3\rangle\approx\frac{1}{\sqrt{2}}(\left|\frac{3}{2}\right\rangle\pm\left|-\frac{3}{2}\right\rangle)$. Using these approximate wave-functions, we note that the hopping matrix element of an excited state $|2\rangle$ to a neighbouring site in its ground state $|0\rangle$ is identical to that in the paramagnetic phase. Therefore, any change in the effective hopping must be because of the small admixture between the two doublets.

\begin{figure}[tb]
\includegraphics[width=0.4\textwidth]{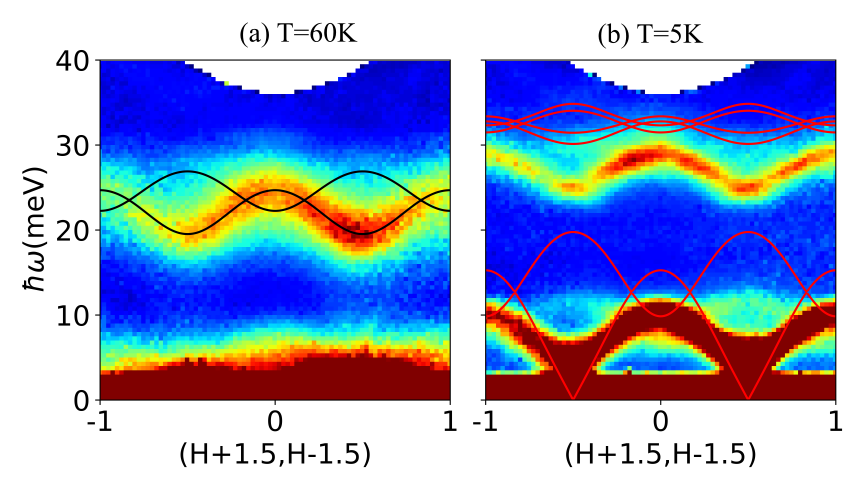}
\caption{INS spectra (a) at 60~K and (b) 5~K above and below the ordering temperature. Solid lines are dispersions calculated using $\mathcal{J}=-~\mathrm{2meV}$ and $\Delta_0=23.5~\mathrm{meV}$. These parameters are chosen so that they reproduce the dispersion in the paramagnetic phase.}
\label{dispersion}
\end{figure}

Quantitatively however, our model considering only a bilinear exchange interaction underestimates the magnitude of bandwidth for the SO exciton by a factor of two. The SO exciton's bandwidths are calculated to be $\sim$2.7~meV and $\sim$3.6~meV for T=0 and $\mathrm{T>T_N}$ (Fig.~\ref{fig3}c), in contrast to experimentally observed values of $\sim$5.4~meV and $\sim$7.2~meV. One way to reproduce the experimentally observed bandwidth is to use a larger exchange interaction $\mathcal{J}=-2~\mathrm{meV}$ and while keeping $\Delta_0=23.5~\mathrm{meV}$. $\Delta_0=23.5~\mathrm{meV}$ is fixed by average energy of the SO exciton for $\mathrm{T>T_N}$. The calculated dispersion is shown in Fig.~\ref{dispersion}, where we overlay the calculation on top of the data. In Fig.~\ref{dispersion}a, the calculation reproduces the data perfectly, as it should. However, as shown in Fig.~\ref{dispersion}b, calculation using the same set of parameters cannot explain our data at $\mathrm{T=5~K}$ at all. First, the calculated magnon bandwidth $\sim$20~meV is almost twice than the experimentally observed value $\sim$12~meV. Second, center of the SO exciton occurs at much higher energy of $\sim$32.5~meV than the observed value $\sim$27~meV. Third, the transition temperature $\mathrm{T_N}$ is predicted to be 85~K, more than twice the experimental value. These inconsistencies are not independent from each other. In a model where $\mathcal{J}$ is the only interaction between spins, it is constrained by the following quantities: $\mathrm{T_N}$, magnon bandwidth as well as the shift of SO exciton going from ordered to paramagnetic phase. To see this, we first note that $\mathrm{T_N}\sim 3\mathcal{J}S_x^2$ within a molecular field approximation. Since $\left\langle S_x\right\rangle\approx\pm 1$ when projected onto the ground doublet, $\mathrm{T_N}\sim40\mathrm{K}$ forces $\mathcal{J}\sim-1~\mathrm{meV}$. On the other hand, the center of the magnon band, roughly half of the full magnon bandwidth, is given by the splitting of the ground state doublet in the single ion model, or $6\mathcal{J}S_x^2$. Using a magnon bandwidth of 12~meV, we obtain a $J\sim -1~\mathrm{meV}$, in agreement with the estimation using $\mathrm{T_N}$. Lastly, from the schematic level diagram in Fig.~\ref{fig3}a, difference in energy of SO exciton at $\mathrm{T=0}$ and $\mathrm{T>T_N}$ is given approximately by half of the splitting in the ground doublet, or $3\mathcal{J}S_x^2$. Using an experimentally observed shift of $\sim$4~meV, we again obtain $J\sim$-1~meV. However, as we discussed in the main text, $J\sim -1~\mathrm{meV}$ could not quantitatively explain the bandwidth of the SO exciton. In other words, there is an `additional' hopping of the SO exciton not explained by bilinear spin interaction.

One possible resolution to this discrepancy is provided by considering quadrupolar interactions\cite{Santini2009}. Interactions between quadrupolar (or higher order) spin moments have been considered extensively for rare earth systems\cite{Levy1979}. They arise naturally from orbital-dependent super-exchange or spin phonon coupling\citep{Birgeneau1969, Santini2009} in the presence of large spin-orbit coupling. Since the quadrupolar operators (see Ref.~\onlinecite{Santini2009} for definition) have no effect when projected onto a pseudo-spin $\mathrm{S}_\mathrm{eff}=\frac{1}{2}$ ground doublet, they will not affect $\mathrm{T_N}$ or the magnon bandwidth to first order. On the other hand, they could contribute to additional hopping of the local SO exciton that might account for the `missing' bandwidth in our model. It will be interesting to search for signature of quadrupolar interactions in other measurements such as parastriction\cite{Morin1980} and non-linear magnetic susceptibility \cite{Morin1981} in the future.

\section{Conclusions}
In conclusion, we have carried out inelastic neutron scattering study of spin-orbit exciton in CoTiO$_3$, a typical example of transition-metal system with strong spin-orbit coupling. We found a complex temperature dependence of the SO exciton across $\mathrm{T_N}$: a significant softening and an increase in bandwidth at T=60~K($>\mathrm{T_N}$) compared to T=5~K($<\mathrm{T_N}$), as well as appearance of another mode at intermediate temperatures below $\mathrm{T_N}$. The observed temperature dependence is satisfactorily explained using a multi-level theory treating simultaneously both the ground and excited multiplets. Quantitatively however, we found the calculated bandwidth considering only bilinear spin interaction to be too small to explain our data, which suggests the presence of higher order spin interaction.

A strong temperature dependence of higher energy SO exciton across $\mathrm{T_N}$ observed in CoTiO$_3$ directly shows that the ground and excited multiplets are strongly coupled in this material. This is in contradiction to a traditional view that these multiplets are decoupled in a transition-metal system. Instead, our observations are strongly reminiscent of behaviours in the rare-earth magnetic materials. Similarity between magnetism in CoTiO$_3$ and rare-earth materials is further highlighted by the success of a multi-level theory originally developed for the rare-earth systems in explaining our data. Our results can be readily generalized to other transition-metal materials with strong SOC, such as the well-known iridates and ruthenates, whose SO excitations are easily probed by INS or resonant inelastic X-ray scattering (RIXS).

\section{Acknowledgement}
The authors thank Arun Paramekanti and Ilia Khait for fruitful discussions. Work at the University of Toronto was supported by the
Natural Science and Engineering Research Council (NSERC)
of Canada. GJS acknowledges the support provided by MOST-Taiwan under project number 105-2112-M-027-003-MY3. FCC acknowledges funding support from the Ministry of Science and Technology (108-2622-8-002-016 and 108-2112-M-001-049-MY2) and the Ministry of Education  (AI-MAT 108L900903) in Taiwan.
This research used resources at the Spallation Neutron Source, a DOE Office of Science User Facility operated by the Oak Ridge National Laboratory. Use of the MAD beamline at the McMaster Nuclear Reactor is supported by McMaster University and the Canada Foundation for Innovation.

\section{Appendix}
\subsection{Dispersion of SO exciton}
\subsubsection{$\mathrm{T>T_N}$}
To obtain the dispersion of SO exciton in the paramagnetic phase, we apply Eq.~\eqref{Green} to the specific case of a honeycomb lattice, where each site has four states shown in Fig.~\ref{fig3}a. Since honeycomb lattice is a non-Bravais lattice with an A and B site, we need to add an additional label to the Green's functions. For each set of $\alpha,\beta=\{+,-,z\}$ of the spin components, we define the Green's function $G_{\mu \nu}^{\alpha\beta}$, where $\mu,\nu=\{A,B\}$ of the honeycomb lattice. Since Eq.~\eqref{SI1} possesses rotational symmetry around $z$ when $h_0=0$, equations of motion for $G_{\mu \nu}^{\alpha\beta}$ is very simple in the paramagnetic phase and is given by 	
\begin{align}
\begin{split}
G_{AA}^{+-}&=g^{+-}+g^{+-}\mathcal{J}(\vec{q})G_{BA}^{+-}\\
G_{BA}^{+-}&=g^{+-}\mathcal{J}(\vec{q})^\star G_{AA}^{+-}.
\end{split}
\label{Green1}
\end{align}
In Eq.~\eqref{Green1}
\footnotesize\begin{align}
\begin{split}
\mathcal{J}(\vec{q})&=\frac{1}{2}\mathcal{J}(1+\exp(-i\vec{q}\cdot\vec{a})+\exp(i\vec{q}\cdot\vec{b}))\\
g^{+-}&=g^{-+}\\
&=\left|\left\langle\frac{3}{2}\left|\tilde{S}^+\right|\frac{1}{2}\right\rangle\right|^2\left[\frac{1}{\omega-\Delta_0}(f_{0}-f_{1})+\frac{1}{\omega+\Delta_0}(f_{1}-f_{0})\right],
\end{split}
\label{symbol1}
\end{align}\normalsize
where $f_{1}$ and $f_{0}$ are the thermal population of a state in the excited doublet $|\pm\frac{3}{2}\rangle$ and ground doublet $|\pm\frac{1}{2}\rangle$, respectively. At $\mathrm{T_N<T\ll \Delta_0}$,$f_{1}\approx0$ and $f_{0}\approx1/2$. $\vec{a}$ and $\vec{b}$ are the in-plane lattice vectors of CoTiO$_3$ crystal structure. $G_{BB}^{+-}$ and $G_{AB}^{+-}$ are related by exactly the same set of equations as Eq.~\eqref{Green1}.  The Green's function $G_{\mu \nu}^{-+}$ are obtained simply by changing $+-\rightarrow -+$ in Eq.~\eqref{Green1}. The poles of Eq.~\eqref{Green1} occur at $1-|g^{+-}\mathcal{J}(\vec{q})|^2=0$, from which we can determine the dispersion of the SO exciton $\omega(\vec{q})=\sqrt{\Delta_0^2\pm \Delta_0\left|\left\langle\frac{3}{2}\left|\tilde{S}^+\right|\frac{1}{2}\right\rangle\right|^2|\mathcal{J}(\vec{q})|}$

We found the longitudinal component of the Green's function, $G^{zz}_{\mu\nu}=0$, as the matrix element $\left\langle J_z=\pm\frac{3}{2}\left|\tilde{S}_z\right| J_z=\pm\frac{1}{2}\right\rangle$ strictly vanishes. Interestingly, this implies the SO exciton observed at $\sim$23~meV in the paramagnetic phase to be purely transverse. This can be easily verified using polarized inelastic neutron scattering in future experiment.

\subsubsection{$T=0$}
Since rotational symmetry about $z$ of Eq.~\eqref{SI} is lost when $h_0\neq 0$, the general form of equations of motion at an intermediate temperature $\mathrm{0< T< T_N}$ is therefore quite complicated. Fortunately, we can easily determine the dispersion at $\mathrm{T=0}$. In this case, only the ground state is populated and only transition from the ground state to an excited state $C_m^\dagger C_0$ is relevant. We can therefore treat the ground state as a vacuum and define a new boson operator, which creates an excited state $|m\rangle$ on an atom $a_m^\dagger\equiv C_m^\dagger C_0$ (m=1,2,3 are the three excited states in Fig.~\ref{fig3}a for $\mathrm{T<T_N}$). For an in-plane ordering along $x$, we can write the spin operators in terms of the newly defined boson operators as\cite{Buyers_1971,Holden_1971}:
\begin{align}
\begin{split}
\tilde{S}_x=\tilde{S}^x_{0,0}&+\sum_m \tilde{S}^x_{m,0}(a_m^\dagger+a_m)\\
&+\sum_{m} (\tilde{S}^x_{m,m}-\tilde{S}^x_{0,0})a_m^\dagger a_m,
\end{split}
\label{Sxordered}
\end{align}
and
\begin{align}
\tilde{S}_+\equiv \tilde{S}_y+i \tilde{S}_z=\sum_m \tilde{S}^+_{m,0} a^\dagger_m+\tilde{S}^+_{0,m} a_m\label{Splusordered}
\end{align}
similarly for $\tilde{S}_-$. It is important to note that the matrix element $\tilde{S}^\alpha_{m,0}\equiv\left\langle m\left| \tilde{S}^\alpha\right| 0\right\rangle$ is evaluated between the basis functions in the ordered phase, where a non-zero molecular field is present, and NOT the basis functions given by Eq.~\eqref{W1} and Eq.~\eqref{W2}. The matrix elements $\tilde{S}^\alpha_{m,n}$ in this new basis are given by:
\begin{align}
\tilde{S}_x=\begin{pmatrix}
	1.07 &0 &0 &-0.68\\
	0&-0.74&  0.87& 0\\
	0& 0.87& -0.19& 0\\
	-0.68& 0& 0& -0.13
\end{pmatrix},\label{Sxmatrixordered}
\end{align}

and

\begin{align}
\tilde{S}_+=\begin{pmatrix}
	0& -1.55i& -0.72i &0\\
	0.35i & 0 & 0 & 0.97i\\
	0.61i & 0 & 0 & 1.33i\\
	0 & -0.76i & 1.39i & 0
\end{pmatrix}.\label{Splusorderedmatrix}
\end{align}
For each matrix $\tilde{S}^\alpha$ defined above, we define the following column vectors $\vec{u}_\alpha=(\tilde{S}^{\alpha}_{10},\tilde{S}^{\alpha}_{20},\tilde{S}^{\alpha}_{30})^\mathsf{T}$ and $\vec{v}_\alpha=(\tilde{S}^{\alpha}_{01},\tilde{S}^{\alpha}_{02},\tilde{S}^{\alpha}_{03})^\mathsf{T}$ which will be used later.

Using the $a_m^\dagger$-boson representations of the spin operators, the total Hamiltonian, which is a sum of the single-ion and the interaction part, $H_1+H_2$ can be reformulated as a quadratic boson Hamiltonian that is easily diagonalized by the standard Bogoliubov transformation.

Since the state $|m\rangle$ created by $a_m^\dagger$ already diagonalizes $H_1$. $H_1$ in terms of boson operators are simply:
\begin{align}
H_1=\sum_{m, i}\hbar\omega_m(a_{m,i}^\dagger a_{m,i}+b_{m,i}^\dagger b_{m,i}),\label{Hsi}
\end{align}
where the index $m$ runs over the three excited states (1,2,3) in Fig.~\ref{fig3}a and index $i$ runs over all unit cells. To distinguish between the two sub-lattices A and B in each unit cell, bosons created on A and B are denoted by $a^\dagger$ and $b^\dagger$ respectively.

Moving on to the interaction part, $H_2$. In spin operators, it is given by:
\begin{align}
\begin{split}
H_2=&\mathcal{J}\sum_{i,\delta}\tilde{\vec{S}}(A,i)\cdot\tilde{\vec{S}}(B,i+\delta)\\
-&\mathcal{J}\sum_{i}[3\left\langle \tilde{S}^x\right\rangle (\tilde{S}^{x}(A,i)+\tilde{S}^{x}(B,i))].\end{split}\label{Hi}
\end{align}

In this expression, $\tilde{S}(A,i)$ denotes the spin on sub-lattice A of unit cell i. The first term, which sums over all unit cell i and three nearest neighbour $\delta$ of each unit cell, gives the Heisenberg interaction between nearest neighbours. The second term is the mean-field part of the first term. It has to be removed as it is already included in $H_1$. Written in terms of boson operators $\psi_{A,i}=(a_{1,i}^\dagger,a_{2,i}^\dagger,a_{3,i}^\dagger,a_{1,i},a_{2,i},a_{3,i})^\mathsf{T}$ (Similarly for the sub-lattice B), it is given by:
\begin{align}
H_2=\mathcal{J}\sum_{i,\delta}\psi_{A,i}^\mathsf{T}\mathsf{H}\psi_{B,i+\delta},\label{Hiboson}
\end{align}
where column vectors $\vec{u}_\alpha$ and $\vec{v}_\alpha$ are defined above. In this expression, $\mathsf{H}$ is a real symmetric matrix given by:
\footnotesize\begin{align}
\mathsf{H}=\mathcal{J}\left[\begin{pmatrix}
\vec{u}_x\\
\vec{v}_x
\end{pmatrix}
\begin{pmatrix}
\vec{u}_x^\mathsf{T}&\vec{v}_x^\mathsf{T}
\end{pmatrix}+\frac{1}{2}\begin{pmatrix}
\vec{u}_+\\
\vec{v}_+
\end{pmatrix}
\begin{pmatrix}
\vec{u}_-^\mathsf{T}&\vec{v}_-^\mathsf{T}
\end{pmatrix}+\frac{1}{2}\begin{pmatrix}
\vec{u}_-\\
\vec{v}_-
\end{pmatrix}
\begin{pmatrix}
\vec{u}_+^\mathsf{T}&\vec{v}_+^\mathsf{T}
\end{pmatrix}\right].\label{Himatrix}
\end{align}\normalsize
$\mathsf{H}$ consists of four symmetric 3 by 3 sub-matrices. As an example, the sub-matrix $\mathsf{H}_{11}=\mathcal{J}[\vec{u}_x\vec{u}_x^\mathsf{T}+\frac{1}{2}(\vec{u}_+\vec{u}_-^\mathsf{T}+\vec{u}_-\vec{u}_+^\mathsf{T})]$

Going to momentum space, and defining the column vector of operators $\psi_k=(a_{k,1},..,a_{k,3},b_{k,1},..,b_{k,3},a_{-k,1}^\dagger,..,a_{-k,3}^\dagger,b_{-k,1}^\dagger,..,b_{-k,3}^\dagger)^\mathsf{T}$, the total Hamiltonian is given by:
\begin{align}
H_1+H_2=\frac{1}{2}\sum_k\psi_k^\dagger \mathsf{H}_k\psi_k,\label{Hk}
\end{align}
where
\begin{align}
\mathsf{H}_k=\begin{pmatrix}
	\mathsf{D}& \mathsf{H}_{12}\gamma(\vec{k})&0&\mathsf{H}_{11}\gamma(\vec{k})\\
	\mathsf{H}_{12}\bar{\gamma}(\vec{k}) & \mathsf{D} & \mathsf{H}_{11}\bar{\gamma}(\vec{k}) & 0\\
	0 & \mathsf{H}_{22}\gamma(\vec{k}) & \mathsf{D} & \mathsf{H}_{21}\gamma(\vec{k})\\
	\mathsf{H}_{22}\bar{\gamma}(\vec{k}) & 0 & \mathsf{H}_{21}\bar{\gamma}(\vec{k}) & \mathsf{D}
\end{pmatrix},\label{hk}
\end{align}
where $\gamma(\vec{k})=(1+\exp(-i\vec{q}\cdot\vec{a})+\exp(i\vec{q}\cdot\vec{b}))$. In Eq.~\eqref{hk},$\mathsf{H}_{ij}$ is the 3 by 3 sub-matrix of $\mathsf{H}$ defined above and $\mathsf{D}$ is a 3 by 3 diagonal matrix with entries $(\hbar\omega_1,\hbar\omega_2,\hbar\omega_3)$ along the diagonal. Eq.~\eqref{Hk} can be diagonalized by finding the positive eigenvalues of the non-Hermitian matrix $\mathsf{g}\mathsf{H}_k$ where $\mathsf{g}$ is a 12 by 12 diagonal matrix with the first six entries given by +1 and last six entries given by -1. Dispersion for $\mathrm{T>T_N}$ and $\mathrm{T=0}$ are shown in Fig.~\ref{fig3}c.

\end{document}